\documentclass{INTERSPEECH2023}
\usepackage{caption}
\usepackage{subcaption}
\usepackage{comment}

\newcommand{\Lagr}{\mathcal{L}}
\newcommand\norm[1]{\left\lVert#1\right\rVert}
\interspeechcameraready

\title{Investigating self-supervised features for expressive, \\ multilingual voice conversion}
\name{Álvaro Martín-Cortinas, Daniel Sáez-Trigueros, Grzegorz Beringer, Iván Vallés-Pérez, Roberto Barra-Chicote, Biel Tura-Vecino, Adam Gabryś, Piotr Bilinski, Thomas Merritt, Jaime Lorenzo-Trueba}
\address{Amazon AGI}
\email{\{alvaromn, beringg, ivallesp, rchicote, bieltura\}@amazon.com}

\begin{document}

\maketitle
 
\begin{abstract}
Voice conversion (VC) systems are widely used for several applications, from speaker anonymisation to personalised speech synthesis. Supervised approaches learn a mapping between different speakers using parallel data, which is expensive to produce. Unsupervised approaches are typically trained to reconstruct the input signal, which is composed of the content and the speaker information. Disentangling these components is a challenge and often leads to speaker leakage or prosodic information removal. In this paper, we explore voice conversion by leveraging the potential of self-supervised learning (SSL). A combination of the latent representations of SSL models, concatenated with speaker embeddings, is fed to a vocoder which is trained to reconstruct the input. Zero-shot voice conversion results show that this approach allows to keep the prosody and content of the source speaker while matching the speaker similarity of a VC system based on phonetic posteriorgrams (PPGs).
\end{abstract}

\section{Introduction}
The goal of many-to-many voice conversion (VC) systems is to change the identity of a source speaker to that of a target speaker. Depending on the learning paradigm applied and the data needed for the training process, these systems can be divided in two \cite{sisman20-vcoverview}: supervised (trained with parallel data) and unsupervised (trained with non-parallel data). 

Supervised voice conversion systems require parallel data~\cite{zhang19-seq2seqvc, serrano19-par-nonpar-vc, zhang20-deepconversion}, i.e., pairs of recordings with the same content but different speaker identities, which are often expensive to produce~\cite{chen22-dataaug-nonparallelvc}. On the other hand, unsupervised voice conversion systems do not require non-parallel data, but have the additional difficulty of establishing a mapping between the source and target utterances~\cite{sisman20-vcoverview}. 

Typically, non-parallel VC systems make use of carefully designed bottlenecks~\cite{qian19-autovc}, variational autoencoders~\cite{saito18-vae}, generative adversarial networks (GANs)~\cite{kaneko19-cycleganvc2, kameoka18-starganvc}, normalizing flows \cite{merritt2022-textfree-nfvc} or intermediate representations such as text or PPGs extracted from automatic speech recognition (ASR) models~\mbox{\cite{sisman20-vcoverview,ezzine22-nonparallelvc-autoregressive}}. Nevertheless, GANs are known to be difficult to train \cite{qian19-autovc}, normalizing flows are text-conditioned or have a lower performance when the flow prior is trained jointly with its weights \cite{merritt2022-textfree-nfvc}, bottlenecks usually drop desired information or suffer from speaker leakage even when carefully designed~\cite{qian19-autovc}, and PPGs require external prosody information to control the prosody of the converted speech~\cite{lian21-finegrained-prosody}. 

More recently, non-parallel VC models based on self-supervised speech representations (S3Rs) as intermediate features~\cite{lin21-fragmentvc, huang22-s3rvc} have been proposed. One important advantage of using S3Rs instead of PPGs as intermediate representations is that self-supervised learning (SSL) models are trained in a completely unsupervised way, whereas ASR models need curated datasets of audio and its corresponding transcriptions. However, when using S3Rs the content and speaker information must be disentangled, in contrast with PPGs which are assumed to be speaker-independent. 

Disentanglement of S3R features is usually performed by applying a bottleneck~\cite{lin21-fragmentvc} or discretizing them~\cite{huang22-s3rvc}. Nevertheless, these techniques have several drawbacks. First, in~\cite{lin21-fragmentvc} the bottleneck forces the removal of information from the source speech, which may not be only speaker information. Second, in~\cite{huang22-s3rvc} the discretization techniques resulted in poor intelligibility even with large codebooks. Finally, in both approaches only the last layer of all SSL models is used, even though the different layers of SSL models have been proved to contain different information~\cite{chen22-wavlm, pasad2021-layerwisessl, polyak21-resynthesis-s3r, li22-ssl-layers}. In particular, for prosody-intensive tasks, such as expressive VC, a weighted-sum of the layers of SSL models has proven to give good results~\cite{lin22-prosody-ssl}.

Thus, this paper describes the following contributions:
\begin{itemize}
    \item We propose a methodology for extracting content features for voice conversion as an average of carefully selected layers of WavLM~\cite{chen22-wavlm} and HuBERT~\cite{hsu21-hubert}, without any previous quantization so that all the content information is kept.
    \item We propose \textit{Chameleon}, a method to automatically extract content representations with a learnable weighted average of the hidden states of the SSL model.
    \item We show that S3Rs trained only with English data achieve a higher intelligibility and preserve the prosody better in zero-shot multilingual VC than PPGs, even with PPGs extracted from a multilingual ASR.
\end{itemize}

\section{Methods}
\begin{figure}[ht]
  \centering
  \includegraphics[width=0.9\linewidth, trim={5cm 24.2cm 4.5cm 1cm}, clip]{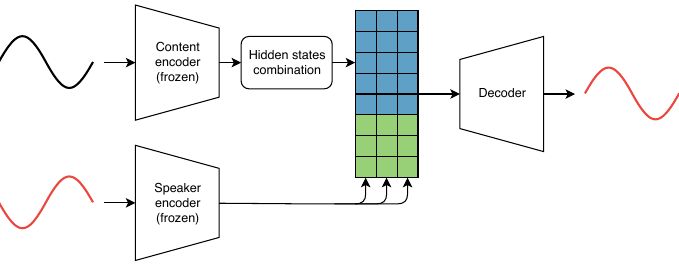}
  \caption{Voice conversion general architecture at inference.}
  \label{fig:general-arch-inference}
\end{figure}

As illustrated in Figure \ref{fig:general-arch-inference}, all VC systems in this paper are built with a similar architecture: a content encoder whose hidden layer's outputs are combined, a speaker encoder and a decoder. At training, the models reconstruct the input signal from the content features concatenated with the speaker embedding, with the assumption that the content features do not contain any speaker information. At inference, the model performs voice conversion by taking the speaker information from an utterance of the target speaker, and the content information from an utterance of the source speaker. 

\begin{comment}
\begin{figure}[t]
    \centering
    \begin{subfigure}{\linewidth}
      \centering
      \includegraphics[width=0.8\linewidth, trim={0 24.5cm 8cm 0.2cm}, clip]{img/general-arch-training.pdf}
      \caption{Voice conversion general architecture at training.}
      \label{fig:general-arch-training}
    \end{subfigure}\hfill
    \begin{subfigure}{\linewidth}
      \centering
      \includegraphics[width=0.8\linewidth, trim={5cm 24.2cm 4.5cm 1cm}, clip]{img/general-arch-inference.pdf}
      \caption{Voice conversion general architecture at inference time.}
      \label{fig:general-arch-inference}
    \end{subfigure}
    \caption{Voice conversion general architecture.}
    \label{fig:general-arch}
\end{figure}
\end{comment}

For the decoder, the universal vocoder BigVGAN~\cite{lee2023-bigvgan} has been selected for its capabilities to generalize to unseen speakers and languages. BigVGAN is trained with the loss computed for the generator and the discriminator as described in~\cite{lee2023-bigvgan}.

As for the speaker encoder, two pre-trained and frozen models have been tested. The first is a speaker verification model based on the Generalized End-to-End Loss (GE2E-Loss embeddings) for its decreased training time and speaker verification Equal Error Rate~\cite{wan2018-ge2e}, which was trained with proprietary data and Common Voice~\cite{ardila2019-commonvoice}. For the second speaker encoder, a pre-trained version of WavLM with an X-vector head~\cite{snyder2018-xvectors} has been used, which allows to extract both the content information from WavLM hidden states and the speaker information from the X-vector head.

Content features are extracted using one of three pre-trained and frozen models: Whisper\footnote{Whisper: \url{https://openai.com/blog/whisper/}.}, WavLM~\cite{chen22-wavlm} and HuBERT~\cite{hsu21-hubert}. Whisper's intermediate features (PPGs) can be used to extract precise phonetic content without speaker information, i.e. requiring no additional disentanglement. Regarding WavLM and HuBERT, although WavLM is the state-of-the-art in most downstream tasks according to the SUPERB benchmark~\cite{yang2021-superb}, HuBERT has also been included in this study to test the architecture proposed with more than one SSL model.

In the following subsections, the different methodologies proposed to generate the content features from the hidden representations of the content encoders are described in detail.

\subsection{Whisper} \label{subsec:whisper}
Various VC systems based on phonetic posteriorgrams (PPGs) extracted from ASR systems can be found in the literature~\cite{sun16-ppgs, serrano19-ppgs, chen20-ppgs}. In this paper we train baseline VC models for English and multilingual setups using PPGs extracted using Whisper as ASR. Whisper is selected to obtain these intermediate representations as it is a multilingual ASR that has proven to be robust across various datasets and languages.

Whisper is built upon a transformer architecture~\cite{vaswani2017-transformer}, where  a latent representation of the spectrogram is derived by the transformer encoder. Thus, in the models trained with Whisper as content encoder, the hidden states of the last layer of its transformer encoder are used as content features, which would be equivalent to PPGs or bottleneck features (BNFs)~\cite{huang22-s3rvc}. These features are not expected to contain speaker information, as it is not needed for the ASR task, hence no further disentanglement is needed. 

For the monolingual case, only Whisper base (74M parameters) is considered. For the multilingual case, both Whisper-base and Whisper large v2 (1550M parameters) are considered to check if there are any significant differences in performance for languages where less training data was available. 

\subsection{WavLM and HuBERT with a fixed average} \label{subsec:fixed-average}
The authors of~\cite{chen22-wavlm, pasad2021-layerwisessl} analyse and report the importance of each layer in different SSL models for different downstream tasks. Intuitively, the results show that higher layers are more related to abstract concepts, such as words, whereas lower layers are more related to local signal properties and low-level speech characteristics, such as speaker identity.

Consequently, useful content features should be extracted from hidden states of the last layers, since first layers most likely contain the most speaker information. For that reason, a fixed average of layers 8 to 12 in WavLM base+, and 7 to 12 in HuBERT base, have been used as a first approach to obtain the content features. These particular layers have been selected according to Figure 2 in~\cite{chen22-wavlm}.

\subsection{WavLM with a learned weighted average (Chameleon)} \label{subsec:chameleon}
Carefully selecting layers in self-supervised learning models is a costly method, as it requires to evaluate the importance of each layer in different downstream tasks and trying different combinations of them to find the best configuration. Thus, learning which hidden states are the most important for the task at hand is a more scalable and general approach.

We propose a novel model, Chameleon, that generates the content features from the self-supervised hidden states by learning, per dimension, the linear combination of the layers of the SSL model that minimizes the decoder loss, and maximizes the disentanglement with the speaker embeddings. 
%With this approach, it is assumed that the features can be combined independently. The reason behind that assumption is that the combination of the features is already done in the feed-forward neural networks of the SSL model if it follows a transformer architecture, as WavLM and HuBERT. Furthermore, making this assumption greatly reduces the number of parameters needed.

\begin{figure}[ht]
  \centering
  \includegraphics[width=\linewidth, trim={0 22.75cm 4cm 0.2cm}, clip]{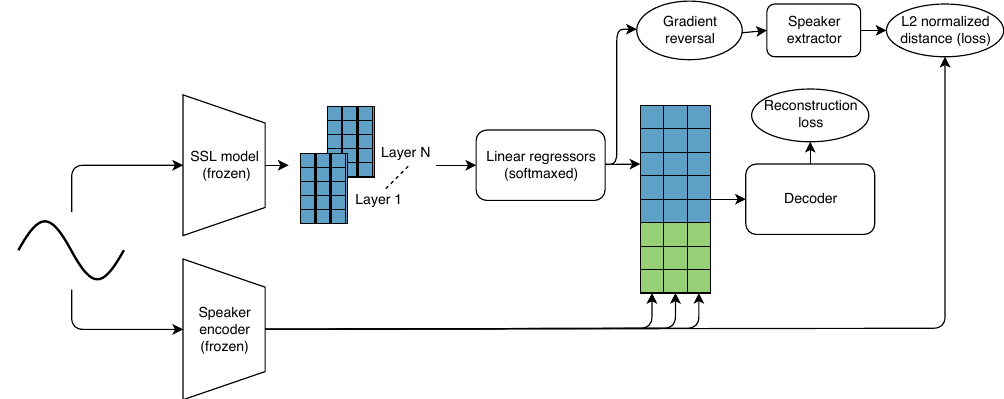}
  \caption{Chameleon architecture during training.}
  \label{fig:chameleon-paper}
\end{figure}

To enforce the disentanglement, a L2 distance adversarial loss with gradient reversal~\cite{schnell21-emocat} is added between the pre-trained speaker embedding and a predicted speaker embedding conditioned with the content features. Intuitively, to maximize the L2 distance the model will learn a weighted average of hidden states that cannot be used to predict the pre-trained speaker embedding. Figure \ref{fig:chameleon-paper} illustrates the architecture of Chameleon during training. At inference, the L2 distance is no longer needed. 

Mathematically, the loss of the generator \cite{lee2023-bigvgan} is modified to be:
\begin{align}
  \Lagr_G = \Lagr_{BigVGAN} + \lambda_{L2}(w)\norm{\Vec{s} - \Vec{\hat{s}}}^2,
\end{align}
where $\Vec{s}$ is the pre-trained speaker embedding, $\Vec{\hat{s}}$ is the predicted speaker embedding and $\lambda_{L2}(w)$ is the weight of the L2 distance in the generator loss. $\lambda_{L2}(w)$ is a function of the parameter $w$ in consideration, as it is positive for the speaker extractor parameters, negative for the linear regressors (due to the gradient reversal) and zero for the decoder parameters (they do not contribute to the L2 distance).

The speaker extractor that predicts the pre-trained speaker embedding is based on a transformer encoder with a CLS embedding~\cite{devlin19-bert}. No positional embedding is added to prevent the transformer from learning the content. This approach assumes that the speaker identity is independent of the ordering of the frames. After the transformer encoder, the CLS is linearly projected to match the the dimensionality of the speaker embedding.

\section{Experiments and results}
\subsection{Naming convention}
In this section, the trained models are presented with the following naming convention:
\begin{itemize}
    \item \texttt{VCWhisper base} and \texttt{VCWhisper large v2} refer to the VC models trained with Whisper base and large v2 as content encoder (see Section \ref{subsec:whisper}), respectively, and the speaker verification model with GE2E-Loss embeddings.
    \item \texttt{VCWavLM base+} and \texttt{VCWavLMX base+} refer to VC models trained both with WavLM base+ as content encoder with a fixed average of layers 8 to 12 (see Section \ref{subsec:fixed-average}), using as speaker embeddings GE2E-Loss embeddings for the former and x-vectors for the latter.
    \item \texttt{VCHuBERT base} refers to a VC model trained with HuBERT base as content encoder with a fixed average of layers 7 to 12 (see Section \ref{subsec:fixed-average}) and GE2E-Loss embeddings as speaker embeddings.
    \item \texttt{Chameleon} refers to the model described in Section \ref{subsec:chameleon}. In these experiments, WavLM base+ is used as the SSL model, because it is better than HuBERT in SUPERB \cite{yang2021-superb}, and x-vectors are used as speaker embeddings for simplicity because they can be directly extracted from WavLM base+.
\end{itemize}
All models have been trained with 8 NVIDIA Tesla V100 SXM2 16GB GPUs with an average training time of 3 days.

\subsection{Evaluation metrics}
Models are evaluated with both objective and subjective metrics. Subjective evaluations are conducted by 100 testers and 10 submissions per tester to evaluate the speaker similarity and the naturalness of the synthesized speech. Paired t-tests are used to determine whether there are significant differences between the models, which require a corrected p-value using Holm method \cite{holm79-correctedpvalue} below 0.05. Objective evaluations consist of two metrics: the word error rate (WER), as a proxy of the intelligibility, and the F0 correlation between the source and the converted utterances, as a proxy of the extent to which the prosody of the source utterance is kept. 

\subsection{Monolingual experiments}
\noindent\textbf{Experimental setup.} English models have been trained with LibriTTS~\cite{zen19-libritts} train-other-500 subset, which is composed of 310 hours of speech and 1160 speakers. To test English models, the source utterances are extracted from LibriTTS test-other subset, which contains  6.69 hours of speech with 33 different speakers, and the target speakers are extracted from LibriTTS test-clean, which contains 8.56 hours of speech and 39 speakers.

\noindent\textbf{Speech intelligibility.} The transcriptions for the WER computation have been generated using Whisper base, with approximately 4,000 utterances. Table \ref{tab:wer-english} shows that all SSL-based models achieve a lower WER than the baseline created with Whisper, and in particular the model based on WavLM with the x-vector head achieves the lowest WER.

\begin{table}[th]
  \caption{WER for English-only models with LibriTTS test-other.}
  \label{tab:wer-english}
  \centering
  \begin{tabular}{ r r }
    \toprule
    \multicolumn{1}{c}{\textbf{Utterances}} & \multicolumn{1}{c}{\textbf{WER}} \\
    \midrule
    Source audio (no VC) & 11.5\%~~~ \\
    VCWhisper base & 14.9\% (+3.4\%)~~~             \\
    VCHuBERT base & 13.6\% (+2.1\%)~~~               \\
    VCWavLM base+ & 13\% (+1.5\%)~~~       \\
    VCWavLMX base+ & \textbf{12.2\% (+0.7\%)}~~~              \\
    Chameleon & 12.4\% (+0.9\%)~~~              \\
    \bottomrule
  \end{tabular}
\end{table}

\noindent\textbf{Prosody.} Table \ref{tab:f0-corr-english} illustrates the F0 correlation with the source utterance for each model. The results show that SSL-based models keep the prosody information of the source utterance better than the model based on Whisper.
%Furthermore, there is no significant difference in terms of prosody with respect to the speaker embeddings used, as VCWavLM base+ and VCWavLMX base+ have the same correlation and they are trained with GE2E speaker embeddings and x-vectors respectively.

\begin{table}[th]
  \caption{F0 correlation to source utterance per monolingual model with 95\% confidence intervals.}
  \label{tab:f0-corr-english}
  \centering
  \begin{tabular}{ r r }
    \toprule
    \multicolumn{1}{c}{\textbf{Model}} & \multicolumn{1}{c}{\textbf{F0 correlation}} \\
    \midrule
    VCWhisper base & $59.3 \pm 0.9$~~~ \\
    VCHuBERT base & \bm{$67.3 \pm 0.9$}~~~             \\
    VCWavLM base+ & \bm{$65.7 \pm 1.0$}~~~               \\
    VCWavLMX base+ & $64.7 \pm 1.0$~~~               \\
    Chameleon & $62.9 \pm   0.9$~~~       \\
    \bottomrule
  \end{tabular}
\end{table}

\begin{comment}
\begin{figure}[h]
  \centering
  \includegraphics[width=0.8\linewidth]{img/f0-corr-english.png}
  \caption{F0 correlation to source utterance per monolingual model.}
  \label{fig:f0-corr-english}
\end{figure}
\end{comment}

\noindent\textbf{MUSHRA speaker similarity and naturalness.} The tests are composed of 100 testcases, where 15\% are male-to-female, 15\% female-to-male, 30\% are the utterances with the highest WER, and the rest are randomly chosen. Cross-gender examples are included because they are the most difficult for speaker similarity, whereas the samples with the highest WER are expected to be the most complicated in terms of intelligibility. Table \ref{tab:subjective-english-ge2e} illustrates that SSL-based models achieve the same naturalness and speaker similarity as the baseline Whisper.

\begin{table}[th]
  \caption{Naturalness and speaker similarity with 95\% confidence intervals for English models trained with GE2E-Loss embeddings.}
  \label{tab:subjective-english-ge2e}
  \centering
  \begin{tabular}{ r r r }
    \toprule
    \multicolumn{1}{c}{\textbf{Utterances}} & \multicolumn{1}{c}{\textbf{Naturalness}} &
    \multicolumn{1}{c}{\textbf{Speaker similarity}}\\
    \midrule
    Ground Truth (GT) & $72.2 \pm 1.4$ & $76.6 \pm 1.5$~~~ \\
    VCWhisper base & $69.9 \pm 1.5$ & $73.5 \pm 1.7$~~~             \\
    VCHuBERT base & $70.0 \pm 1.4$ & $72.4 \pm 1.7$~~~               \\
    VCWavLM base+ & $70.4 \pm 1.5$ & $73.2 \pm 1.7$~~~       \\
    \bottomrule
  \end{tabular}
\end{table}

\begin{table*}[t]
  \caption{WER per language for multilingual models with LibriTTS test-other.}
  \label{tab:wer-multilingual}
  \centering
  \begin{tabular}{ r r r r r r r r r }
    \toprule
    \multicolumn{1}{c}{\textbf{Utterances}} & 
    \multicolumn{1}{c}{\textbf{Dutch}} &
    \multicolumn{1}{c}{\textbf{English}} &
    \multicolumn{1}{c}{\textbf{French}} &
    \multicolumn{1}{c}{\textbf{German}} &
    \multicolumn{1}{c}{\textbf{Italian}} &
    \multicolumn{1}{c}{\textbf{Polish}} &
    \multicolumn{1}{c}{\textbf{Portuguese}} &
    \multicolumn{1}{c}{\textbf{Spanish}} \\
    \midrule
    Source audio (no VC) & 9.2\% & 7.7\% & 7.8\% & 6.3\% & 13.3\% & 5.6\% & 8.4\% & 5.4\%~~~ \\
    VCWhisper base & 11.5\% & 9.4\% & 9.2\% & 8.6\% & 15.8\% & 7.9\% & 10.9\% & 6.0\%~~~             \\
    VCWhisper large v2 & 11.4\% & 9.7\% & 9.1\% & 8.5\% & 16.0\% & 7.6\% & 10.3\% & 6.7\%~~~             \\
    VCHuBERT base & 10.1\% & 8.4\% & 8.6\% & 7.4\% & 14.8\% & 6.5\% & 9.8\% & 6.1\%~~~               \\
    VCWavLMX base+ & 10.0\% & 8.5\% & \textbf{8.4\%} & 7.1\% & 14.6\% & \textbf{6.3\%} & 10.1\% & \textbf{5.4\%}~~~       \\
    Chameleon & \textbf{9.6\%} & \textbf{8.3}\% & 8.6\% & \textbf{6.8\%} & \textbf{14.4\%} & \textbf{6.3\%} & \textbf{9.3\%} & 6.0\%~~~\\
    \bottomrule
  \end{tabular}
\end{table*}

%Table \ref{tab:subjective-english-wavlm} also shows that Chameleon successfully learns which hidden states are the most important to represent the content of the source utterance and achieves the same speaker similarity and naturalness as the other models based on WavLM. To illustrate this, Figure \ref{fig:chameleon-weights} shows the softmaxed weights learned by Chameleon, where the highest weights are located in layers 8 to 12 as expected.

\begin{figure}[ht]
  \centering
  \includegraphics[width=\linewidth, trim={0 0.4cm 0 1.3cm}, clip]{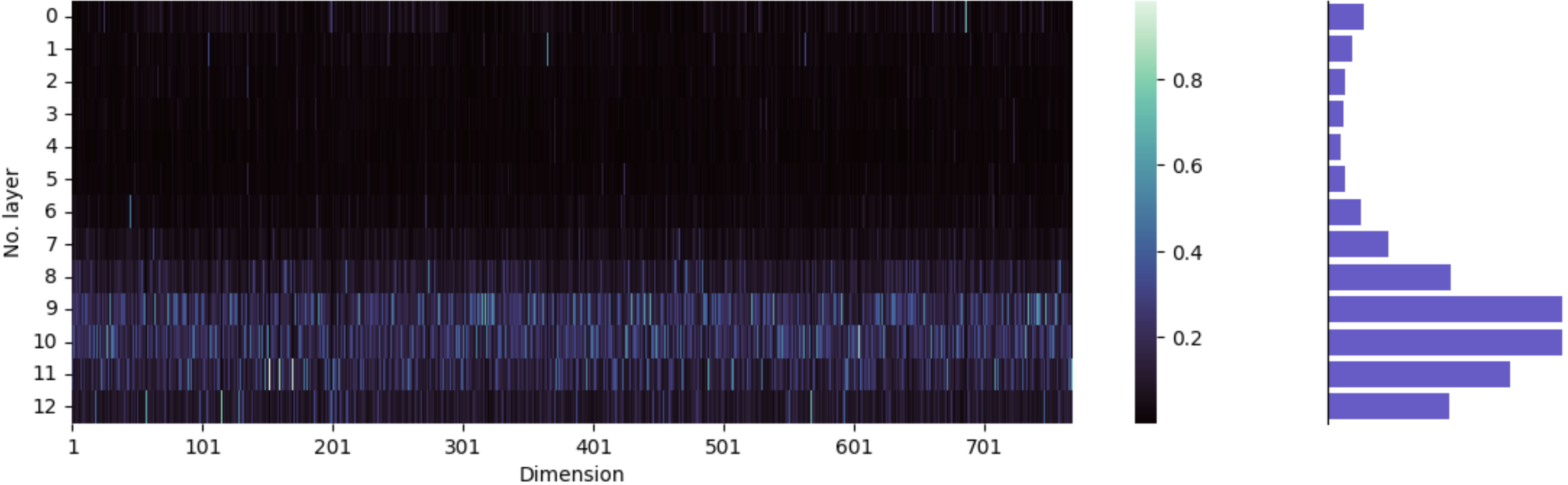}
  \caption{Chameleon weights with the histogram per layer.}
  \label{fig:chameleon-weights}
\end{figure}

Similar tests were conducted to compare Chameleon, which learns the weighted average of WavLM hidden states, with VCWavLM base+ and VCWavLMX base+, which perform a fixed average. The results, which are not included due to space restrictions, show that the three models have no significant difference in terms of speaker similarity and naturalness. Finally, Figure \ref{fig:chameleon-weights} shows that Chameleon has learned to give more weight to the hidden states of layers 8 to 12, the same layers manually selected for the fixed average in VCWavLM base+ and VCWavLMX base+.

\begin{comment}
\begin{table}[th]
  \caption{Subjective evaluations for English models trained with WavLM as content encoder.}
  \label{tab:subjective-english-wavlm}
  \centering
  \begin{tabular}{ r r r }
    \toprule
    \multicolumn{1}{c}{\textbf{Utterances}} & \multicolumn{1}{c}{\textbf{Naturalness}} &
    \multicolumn{1}{c}{\textbf{Speaker similarity}}\\
    \midrule
    GT & $76.3 \pm 1.2$ & $74.7 \pm 1.3$~~~ \\
    VCWavLM base+ & $73.0 \pm 1.4$ & $71.4 \pm 1.5$~~~             \\
    VCWavLMX base+ & $74.0 \pm 1.3$ & $71.1 \pm 1.5$~~~               \\
    Chameleon & $72.4 \pm 1.4$ & $70.6 \pm 1.6$~~~       \\
    \bottomrule
  \end{tabular}
\end{table}
\end{comment}

\subsection{Multilingual experiments}
\noindent\textbf{Experimental setup.} Multilingual models have been trained with a balanced subset of 200,000 utterances from the train split of Multilingual LibriSpeech (MLS)~\cite{pratap20-mls}, where 25,000 were included per language: English, German, Dutch, Spanish, French, Italian, Portuguese and Polish. Validation and test subsets are created from MLS development and test splits respectively, with a total of 4,000 utterances each, 500 utterances per language. For testing, the target speakers are all English speakers from LibriTTS test-clean.

\noindent\textbf{Speech intelligibility.} The transcriptions for the WER computation have been generated using Whisper large v2, as it has a significantly better performance than Whisper base in languages different to English. Table \ref{tab:wer-multilingual} shows that, as in the English-only case, all SSL-based models achieve a lower WER in most locales than the baselines built with Whisper. The results obtained with Chameleon are particularly remarkable, as it is the model with the lowest WER in 6 out of 8 locales. These results illustrate that even though WavLM was trained with English data only, it can successfully generalise to other locales.

\noindent\textbf{Prosody.} Table \ref{tab:f0-corr-multilingual} illustrates the F0 correlation for each model considering all locales. As in the English-only case, the results show that SSL-based models keep the prosody information of the source utterance better than the model based on Whisper.

\begin{table}[th]
  \caption{F0 correlation to source utterance with 95\% confidence intervals per multilingual model.}
  \label{tab:f0-corr-multilingual}
  \centering
  \begin{tabular}{ r r }
    \toprule
    \multicolumn{1}{c}{\textbf{Model}} & \multicolumn{1}{c}{\textbf{F0 correlation}} \\
    \midrule
    VCWhisper base & $56.4 \pm 0.8$~~~ \\
    VCWhisper large v2 & $58.0 \pm 0.8$~~~ \\
    VCHuBERT base & $68.4 \pm 0.7$~~~             \\
    VCWavLMX base+ & $68.3 \pm 0.8$~~~               \\
    Chameleon & \bm{$72.8 \pm   1.1$}~~~       \\
    \bottomrule
  \end{tabular}
\end{table}

\begin{comment}
\begin{figure}[h]
  \centering
  \includegraphics[width=0.8\linewidth]{img/f0-corr-multilingual.png}
  \caption{F0 correlation to source utterance per multilingual model.}
  \label{fig:f0-corr-multilingual}
\end{figure}
\end{comment}

\noindent\textbf{MUSHRA speaker similarity and naturalness.} Similarly to the monolingual case, MUSHRA tests were conducted to evaluate the speaker similarity and naturalness of the converted utterances with the different models. In particular, Chameleon, VCWavLMX base+ and Whisper large v2 are evaluated in English (to compare with the monolingual case), Italian (high WER) and Spanish (low WER). These models are selected for the MUSHRA test because Chameleon and VCWavLMX have the lowest WER in different locales, and Whisper large v2 is significantly better than Whisper, in F0 correlation and WER in most locales.

As in the monolingual case, in English the results show that the proposed systems have no differences in terms of these two characteristics. Nevertheless, both in Spanish and Italian the SSL-based models are significantly better than VCWhisper large v2 in terms of naturalness. In terms of speaker similarity, Chameleon has a significantly lower speaker similarity in Italian than the other models because the gradient reversal weight in the total loss has to be carefully fine-tuned to reach an optimal value, which is not trivial in multilingual datasets. Table \ref{tab:subjective-spanish} shows the results for Italian, and they are omitted for Spanish and English due to space restrictions.

\begin{comment}
\begin{table}[th]
  \caption{Naturalness and speaker similarity for Spanish.}
  \label{tab:subjective-spanish}
  \centering
  \begin{tabular}{ r r r }
    \toprule
    \multicolumn{1}{c}{\textbf{Utterances}} & \multicolumn{1}{c}{\textbf{Naturalness}} &
    \multicolumn{1}{c}{\textbf{Speaker similarity}}\\
    \midrule
    GT & $65.3 \pm 1.5$ & $76.7 \pm 1.6$~~~ \\
    VCWhisper large v2 & $54.0 \pm 1.6$ & \bm{$43.3 \pm 2.0$}~~~             \\
    VCWavLMX base+ & \bm{$58.3 \pm 1.6$} & \bm{$45.1 \pm 1.9$}~~~               \\
    Chameleon & \bm{$57.8 \pm 1.7$} & \bm{$41.6 \pm 2.0$}~~~       \\
    \bottomrule
  \end{tabular}
\end{table}
\end{comment}
\begin{table}[th]
  \caption{Naturalness and speaker similarity with 95\% confidence intervals for Italian.}
  \label{tab:subjective-spanish}
  \centering
  \begin{tabular}{ r r r }
    \toprule
    \multicolumn{1}{c}{\textbf{Utterances}} & \multicolumn{1}{c}{\textbf{Naturalness}} &
    \multicolumn{1}{c}{\textbf{Speaker similarity}}\\
    \midrule
    GT & $65.1 \pm 1.7$ & $75.9 \pm 1.7$~~~ \\
    VCWhisper large v2 & $53.9 \pm 1.7$ & \bm{$35.4 \pm 1.8$}~~~             \\
    VCWavLMX base+ & \bm{$57.1 \pm 1.7$} & \bm{$35.7 \pm 1.8$}~~~               \\
    Chameleon & \bm{$57.3 \pm 1.7$} & $31.8 \pm 1.7$~~~       \\
    \bottomrule
  \end{tabular}
\end{table}

\section{Discussion}
SSL models outperform ASR models as content encoders both in prosody and intelligibility, as well as naturalness for some languages such as Spanish and Italian. At the same time, they provide equal speaker similarity. As SSL models encode most of the information in the source speech, they contain prosodic information which makes the reconstruction task much simpler and improves the final naturalness and intelligibility of the system. In contrast, ASR models only encode the information relevant for the transcription task, so the prosodic information has to be inferred in detriment of intelligibity and naturalness. 

Nevertheless, in SSL models content and speaker information must be disentagled using an appropriate combination of the model's hidden states which discards those more related to speaker identity. If the disentanglement is not done correctly, the speaker information of the source utterance is used during training to reconstruct the input, and at inference that results in speaker leakage.

\section{Conclusions and future directions}
In this paper, we have proposed computing disentangled content features by carefully averaging the hidden states in different layers of SSL models, either with a fixed average or a learnable weighted average, i.e. Chameleon. The main drawback of performing a fixed average is to previously determine which layers of the SSL model are more related to content, e.g. with downstream tasks. Chameleon's learning paradigm automatically determines those layers, but as a downside the gradient reversal has to be fine-tuned to avoid speaker leakage. In future work, speaker embeddings could also be learned from SSL features in an unsupervised framework similar to Chameleon, forcing the model to separate SSL features into content or speaker features.

\begin{comment}
\section{Acknowledgements}

\ifinterspeechfinal
     The INTERSPEECH 2023 organisers
\else
     The authors
\fi
would like to thank ISCA and the organising committees of past INTERSPEECH conferences for their help and for kindly providing the previous version of this template.
\end{comment}

\bibliographystyle{IEEEtran}
\bibliography{mybib}

\end{document}